\RequirePackage{fixltx2e}
\documentclass[aps,pre,preprint,showpacs,showkeys,superscriptaddress,nofootinbib,longbibliography,floatfix,a4paper]{revtex4-1}
\usepackage{amsmath,amssymb,graphicx,bbm,enumerate,theorem}
\usepackage[caption=false]{subfig}
\usepackage[english]{babel}
\usepackage{xcolor}
\usepackage{dcolumn}
\usepackage{mathtools}
\usepackage[section]{placeins}

\begin{document}

\title{Mobility-limited polyarylamine biscarbonate ester (PABC) /[6,6]-phenyl $C_{61}$ butyric acid methyl ester (PCBM) bulk heterojunction photovoltaic device}

\author{Liang-Bih Lin}
\affiliation{Department of Applied Chemistry, National Chiao Tung University, Hsinchu, 30010 Taiwan}
\author{Krishna Balantrapu}
\affiliation{The Dow Chemical Company, Marlborough, MA  01752 USA}
\author{Amanda Preske}
\affiliation{Department of Chemistry, University of Rochester, Rochester,  NY 14620 USA}
\author{Arthur A. Mamiya}
\author{Dem\'etrio A. da Silva Filho}
\affiliation{Instituto de F\'isica, Universidade de Bras\'ilia, Bras\'ilia, DF, 70919-970 Brazil}
\author{George C. Cardoso}
\email{gcc@usp.br}
\affiliation{ Departmento de F\'isica, Universidade de S\~{a}o Paulo, Ribeir\~{a}o Preto  14040-901, Brazil}

\begin{abstract}

Photovoltaic (PV) devices made from blends of a polyarylamine biscarbonate ester (PABC) and [6,6]-phenyl $C_{61}$ butyric acid methyl ester (PCBM) have been fabricated and characterized. PABC is a hole transporting co-polymer prepared from reacting N,N'diphenyl-N,N'bis(3-hydroxyphenyl)1,1;biphenyl(4,4'diamine), diethylene glycol bischloroformate, and triethylemine. By varying the polymer loading in the blend, optimal power conversion efficiency (PCE) of approximately 0.45\% has been achieved for a blend consisting of 25 wt\% PABC, which is an order of magnitude higher than the PCE for a 45 wt\% blend. The optimal ratio is at about 0.44:0.56 molar ratio of the active hole transporting to electron transporting moieties.  Results of mobility studies suggest that blends with higher PABC loading have efficiencies limited by 'hole' transport. Also responsible for the lower efficiency at higher PABC concentrations was optical filtering. The efficiency does not appear to be limited by deep charge trapping. Density functional theory calculations conducted pointed toward a strong localization of the frontier orbitals, which might offer an explanation for the low PABC hole mobility. Calculated frontier orbital energy levels found band gap to be consistent with the $V_{OC}$ measured.
\end{abstract}


\maketitle{Introduction}
In the race to obtain the best performing organic solar materials, hundreds of materials, blends and configurations have been intensively studied by various groups \cite{li2012polymer,li2012molecular,dou2012tandem}. However, the most efficient single layer organic solar cells still have efficiencies typically below 12\% in laboratory conditions and significantly lower efficiencies in pilot production \cite{heeger201425th}. Therefore, there is a continued need for research in all aspects of polymer photovoltaics such as materials chemistry, materials characterization, device physics, process and production technology. In particular, root cause identification of photovoltaic efficiency limitation is very relevant for further development of solar cells technologies. The most well-known organic photovoltaics are polymer-fullerene bulk heterojunction (BHJ) devices \cite{heeger201425th,dennler2009polymer}, where the polymer is the hole-transport material and the fullerene is the electron-transporting material. Among BHJ photovoltaic devices, the most well studied and most successful one, with efficiencies exceeding 5\%, is the P3HT:PCBM blend, where the hole-transporting material is represented by poly(3-hexylthiophene) (P3HT) and the electron-transporting material is represented by 1-3(-methoxycarbonyl)propyl-1-phenyl[6-6]$C_{61}$ (PCBM). A first step when testing a combination of photovoltaic materials is to establish the optimum ratio of the materials for maximum power conversion efficiency. To achieve high power conversion efficiencies it is necessary to maximize the mobility of the charge carriers and to equalize as much as possible the mobility of the electrons in the electron transport material with the mobility of  holes in the hole-transport material \cite{brabec2008organic}. Power conversion efficiency is also highly dependent on the nano-morphology \cite{moule2006effect} that depends on factors such as the solvent type, solvent evaporation rate, temperatures used during the fabrication process and deposition method \cite{gunes2007conjugated,sobkowicz2011effect,chen2012efficient}.
Most organic solar active materials require post-coating thermal and/or vacuum treatments, as well as additional buffer or blocking layers in their structure to prevent leakage or to facilitate photovoltaic effects \cite{rait2007effect,zhao2010metal}. It would be desirable to develop new photovoltaic devices with materials that support longer storage life and can be fabricated in ambient environment. 
From the theoretical standpoint, density functional theory (DFT) has been widely used to calculate, among other electronic properties of interest for photovoltaics, the frontier orbital energy levels\cite{Zhang2007,McCormick2013,Kozycz2012,Ayoub2015,Hu2015} and their spatial distribution \cite{zhou2012rational,Kanai2007,Fu2015}. Despite there being some discussion with respect to the meaning of Kohn-Sham orbitals obtained from DFT, they are thought to be adequate for qualitative studies (see Stowasser \textit{et al} \cite{Stowasser1999} and references therein).
In this study, polyarylamine biscarbonate ester (PABC)  - a hole-transporting material - was blended with PCBM (electron transporting material) in various ratios to produce solution-processed solar cells. We have investigated the effect of PABC:PCBM blend composition on the power conversion efficiency by examining the I-V curve under standard illumination. We have also studied carrier mobility of the materials and blends using the photoconductive time-of-flight (TOF) technique \cite{lin1998transient,lin2000charge}. 
PABC was originally produced for use as a stand-alone hole transporting polymeric material for photoconductors or photoreceptors in xerographic applications\cite{borsenberger1998organic}.  The molar ratio of the hole transporting moiety-arylamine-to that of carbonate ester moiety is about 62:38.  One of the potential advantages of use of PABC is its long shelf life of over ten years, in contrast with shelf life of polythiophenes, such as P3HT, whose typical storage life is less than 18 months. In contrast to most photovoltaic materials that are highly oxygen sensitive, PABC layers can be in principle fabricated in ambient environments. No electron-blocking layer was necessary to reach photovoltaic activity; however to increase PEC by a factor of approximately 3 we have used PEDOT:PSS as an electron-blocking layer in most of the results reported in this paper. The structure of the device studied is Glass/ITO/ /PEDOT:PSS/PABC:PCBM/Al. 
DTF calculations were used to study energy levels, and to explain open circuit voltage $V_{OC}$ and short circuit current $J_{SC}$ for the PABC:PCBM blend. Calculations were checked against analogous calculations using the P3HT:PCBM blend.

\section*{Materials and Methods}

\subsection*{Materials and PABC:PCBM Solution Preparation}
\begin{figure}[ht]
  \begin{tabular}{ccc}
    \includegraphics[width=0.5\columnwidth]{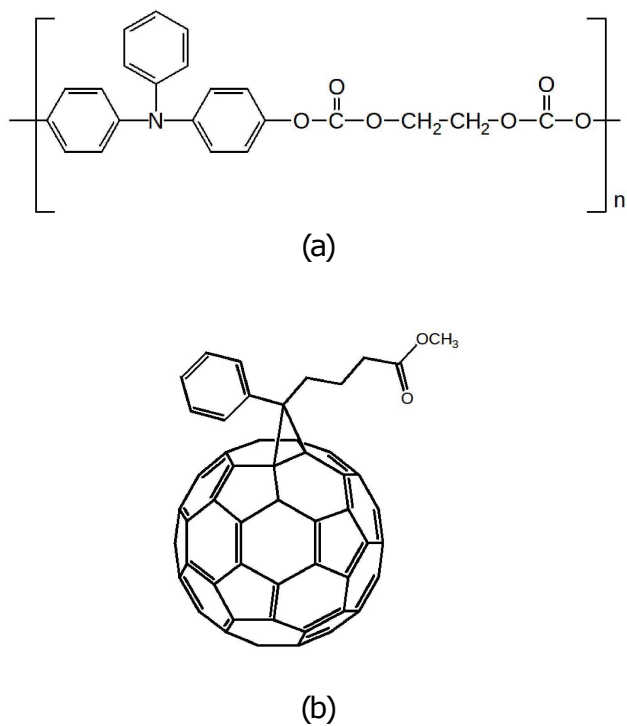}
  \end{tabular}
  \caption{a) PABC molecular structure b) PCBM molecular structure.}
  \label{fgr:fig1}
  \end{figure}

PABC is a co-polymer prepared from reacting N,N'diphenyl-N,N'bis(3hydroxyphenyl)1, 1;biphenyl(4,4' diamine) (DNTBD), diethylene glycol bischloroformate, and triethylemine. The PABC molecular structure is shown in Fig. \ref{fgr:fig1}(a). The PABC was produced by Xerox Corporation. PCBM powder was purchased from American Dye Source. The PCBM molecular structure is shown in Fig. \ref{fgr:fig1}(b). Each material was dissolved in monochlorobenzene and the solution was left overnight. Eight different wt\% of PABC were examined: 5, 15, 20, 25, 30, 35, 40, and 45 wt\%. In this paper, all usage of wt\% refers to the mass percentage of PABC as a fraction of the total mass of solids in the solution, with the remainder of the solid being PCBM. The combined 2 wt\% solutions in the array of PABC:PCBM ratios were mixed in a bath sonicator for 20 minutes and heated to 60$^{\circ}$C for 10 minutes. Each solution was filtered using a 1.0$\mu$m pore-size syringe filter. PEDOT:PSS,  an electrically conductive polymer blend with hole-transporting characteristics that consists of poly(3,4-ethylenedioxythiphene)polycation and a poly(styrenesulfonate)polyanion, was ordered from Sigma Aldrich (150 S/cm at 18 $\mu$m film thickness). Indium-tin-oxide (ITO) coated aluminosilicate glass substrates were purchased from Delta Technologies (1'' x 1'', $\geq$ 85\% transmittance, 120-160 nm coating thickness, 15 Ohms/square).

\subsection*{PABC:PCBM BHJ Solar Cell Fabrication}

\begin{figure}[ht]
  \begin{tabular}{ccc}
    \includegraphics[width=0.5\columnwidth]{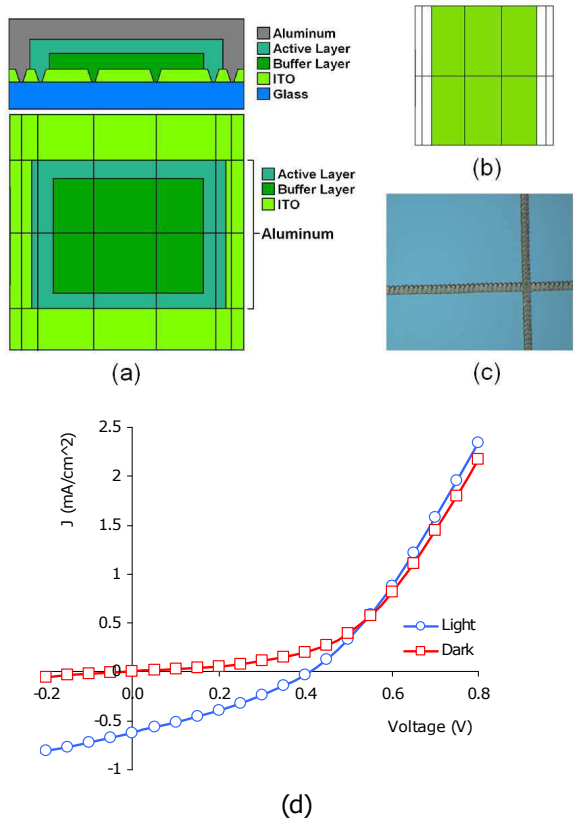}
  \end{tabular}
  \caption{(a) Structure of the photovoltaic cell showing the laser-etched ITO the various deposited layers. The buffer layer used was PEDOT:PSS. (b) Laser etching pattern diagram. (c) Picture of actual laser etched lines. (d) Typical J-V curve observed for one of the six cells shown in (a)}
  \label{fgr:fig2}
  \end{figure}

Instead of the conventional chemical etching the ITO substrates were laser etched to produce six isolated devices. The laser etching produces features under 50 $\mu$m in width and about 0.15 $\mu$m in depth -- enough to completely remove the ITO, as confirmed electrical tests. This is a much faster process than the usual chemical etching of the ITO glass and was readily available in our facility. The laser etching lines as well as the regions where the materials are deposited are shown in Fig. \ref{fgr:fig2}(a)-(c). After laser etching the ITO glass samples were cleaned by ultrasonic bath in isopropyl alcohol.  

Each ITO substrate was etched to form six photovoltaic cells. The area of each cell is approximately 5 mm $\times$ 5 mm. Scotch tape was used to mask the regions not to be coated during the spin coating processes and during the aluminum (cathode) deposition step. PEDOT:PSS was syringe-deposited onto the conveniently masked ITO, spin coated at 4000 rpm for 60 s and air dried to form a layer of about 30 nm thickness. The PABC:PCBM blend solutions were spin-coated at 1100 rpm for 60s onto the ITO coated glass substrates and air dried. The blend solutions were dried in place for 5 minutes to prevent the layer from being disturbed as the volatiles evaporated. The active layer was further annealed at 120$^{\circ}$C for about 15 minutes. The top aluminum contact was deposited by thermal evaporation (Edwards E306A vacuum coating unit) through a mask with aperture of 0.5 cm by 0.5 cm. Aluminum deposition pressure was $3 - 7\times 10^{-6}$ torr . The thickness of the aluminum layer deposited was estimated to be of the order of 100 to 300 nm (barely translucid to room fluorescent light). Deposition time and current were kept constant throughout the various experiments. All processing, except Al deposition was performed at ambient air (RH < 50\%, temperature ~22$^{\circ}$C). During spin coating, a glove box was used to assure that the ambient relative humidity (RH) was below 40\%. Higher relative humidity gave a high rate of very low yield devices.

\subsection*{J-V Characterization}

A solar simulator designed and built by Xerox Corporation and meeting the AM1.5G standard for class B was used as illumination source. The solar simulator spectral characteristics were verified in the 200-1100 nm range using a fiber spectrometer Ocean Optics model HR2000 and the light intensity was calibrated to 100 mW/cm${}^2$ using a thermopile detector Coherent PM3 (50 $\mu$W power resolution). J-V characteristics of dark and illuminated solar cells were measured using a custom-made probe station in conjunction with a Keithley 237 voltage source-measuring unit. J-V data was collected with a Labview routine though the Keithley 237 GPIB. Spring-loaded probes purchased from Everett Charles Technologies were used for electrical contacts with the ITO coated surface region (anode) and with the deposited aluminum region (cathode), both at the same side of the ITO slide. The RC constant of each cell built was of the order of 1 $\mu$s. The scan speed for the J-V curve was chosen such that the dwell time for acquisition of each point was 10 ms or longer to ensure steady state measurements. All measurements were performed within two hours of device fabrication. 

\subsection*{Time-of-flight measurements}

Samples were prepared by bar coating of solution mixtures of the PABC polymer on a pre-coated charge-generating layer consisted of a phthalocyanine on a metalized polyethylene terephthalate substrate. After drying at 120$^{\circ}$C for about 5 minutes and cooling down to room temperature a gold electrode was deposited on the films.  The film thickness was about 5-15 $\mu$m as measured by an eddy current thickness gauge.  Reference devices were also prepared by using mixtures of TPD and polycarbonate (PC) at various weight ratios. TPD is a hole transport molecule N,N'-diphenyl-N,N'-bis(3-methylphenyl)-[1,1'-biphenyl]-4,4'-diamine (TPD) described in reference \cite{khan2004morphological} .Charge mobility was measured in ambient conditions by conventional time-of-flight technique\cite{nam2009high}.

Light pulses were obtained from a nitrogen laser (Spectra Physics) pumped red dye at 680 nm, 10 ns pulse, 10 Hz repetition rate.   The bias of the cells was negative at the Au electrode for hole-transport studies. Reverse biases were also applied to investigate leakage and presence of contaminants.  Data was collected at field strengths of about $1 \times 10^{4}$ to $6 \times 10^{5}$ V/cm. To maintain a constant field during the period of photocurrent transients, the number of injected charges was kept at less than 5\% of the capacitive charges by using neutral density filters in the laser beam. The transit time was determined from the intersection of tangents drawn in linear plots before and after the downturn of the photocurrent transients\cite{abkowitz1998direct}.

\section*{Results and Discussion} 

\begin{figure}[ht]
  \begin{tabular}{ccc}
    \includegraphics[width=0.5\columnwidth]{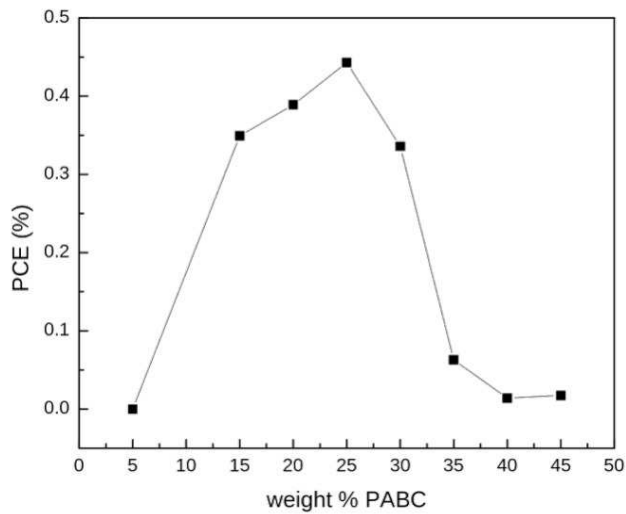}
  \end{tabular}
  \caption{Effect of PABC concentration on power conversion efficiency (PCE)}
  \label{fgr:fig3}
  \end{figure}

The PABC:PCBM solution showed good film forming properties at various weight ratios (5 to 45\% of  PABC).  A typical J-V curve is shown in Fig. \ref{fgr:fig2}(d).  When in dark conditions, the typical diode behavior is shown. For an illumination of 100 mW/cm${}^2$ the photoelectric activity can be observed by the presence of a current density at zero applied voltage. Fig~\ref{fgr:fig3}, shows the power conversion efficiency as a function of the amount of PABC (wt\%).  The highest efficiency achieved without a buffer layer was 0.12\%, for a device having an active layer with 25 wt\% PABC and 75 wt\% PCBM. The efficiency at the same concentrations jumped to 0.45\% with the use of a PEDOT:PSS buffer layer in the position indicated in Fig. \ref{fgr:fig2}(a). When the amount of PABC was increased to 30 wt\%, the efficiency decreased to about 0.32\%.  Further increases in the amount of PABC resulted in devices exhibiting an even lower efficiency: 0.05\% efficiency at 35 wt\% PABC and relatively negligible efficiency at 40 and 45 wt\% PABC. 

Table 1 summarizes the photovoltaic characteristics obtained as a function of PABC concentration, where JSC is the short-circuit current density, Voc is the open circuit voltage,  FF is the fill factor and PCE is the power conversion efficiency.

\begin{table}[!ht]
\centering
\caption{\bf J-V parameters for the different compositions.}
\label{tbl:JVparams}
\begin{tabular}{|l|l|l|l|l|}

\hline
 PABC wt(\%) & JSC(mAcm${}^{-2}$) & V${}_{OC}$(V)& FF &  PCE(\%) \\
 \hline\hline
 
$5$	    &0.00	&0.00	&0.00	&0.00 \\ \hline
$15$	&0.73	&0.24	&0.28	&0.35\\ \hline
$20$	&0.75	&0.26	&0.27	&0.39\\ \hline
$25$	&0.69	&0.32	&0.27	&0.45\\ \hline
$30$	&0.56	&0.30	&0.30	&0.34\\ \hline
$35$	&0.25	&0.13	&0.29	&0.06\\ \hline
$40$	&0.13	&0.06	&0.26	&0.01\\ \hline
$45$	&0.13	&0.07	&0.27	&0.02\\ \hline

\end{tabular}
\end{table}

\begin{figure}[ht]
  \begin{tabular}{ccc}
    \includegraphics[width=0.6\columnwidth]{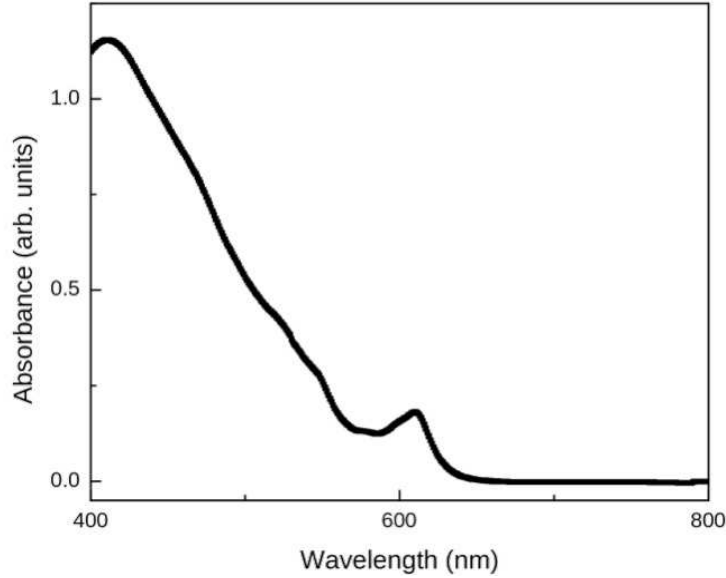}
  \end{tabular}
  \caption{PABC:PCBM absorption spectrum at 25wt\% of PABC.}
  \label{fgr:fig4}
  \end{figure}

We have performed further studies to understand the origins of the efficiency limitation of the PABC:PCBM BHJ solar cell. Fig~\ref{fgr:fig4} shows the absorption spectrum for PABC:PCBM at 25:75. The low efficiency of the PABC/PCBM solar cells can be partially attributed to the predominantly blue visible light absorption of PABC. 

\begin{figure}[ht]
  \begin{tabular}{ccc}
    \includegraphics[width=0.6\columnwidth]{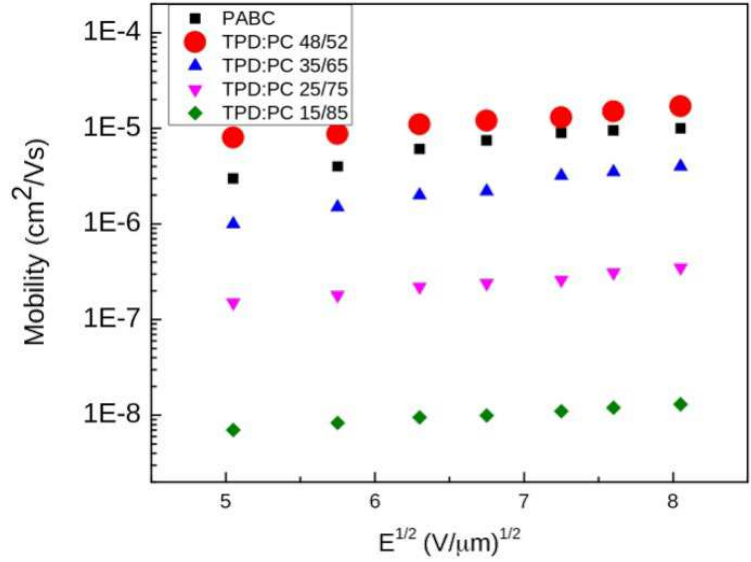}
  \end{tabular}
  \caption{Mobilities measured via time-of-flight for a PABC film and for different percentages by weight of TPD:PC}
  \label{fgr:fig5}
  \end{figure}

In order to understand the role of PABC charge mobility in the determination of the optimum PABC:PCBM ratio and overall PCE, we have performed time of flight experiments. We have used pure PABC films and, for comparison, TPD doped polycarbonate (PC) films, both prepared as previously described. In our photovoltaic studies TPD did not show any appreciable photovoltaic activity when in combination with PCBM at various blending ratios. However, TPD is a well-known hole-transporting  molecule with similar molecular moiety to that of PABC and was used for comparison purposes. Fig~\ref{fgr:fig5} shows the results of our time-of-flight experiments. It was found that the charge transport behaviors of PABC are non-dispersive with a similar field dependency as that of TPD and its charge mobility is comparable to 40-55 wt\% TPD doped PC, suggesting a trap-free hopping charge transport behavior.  The charge mobility of PABC is somewhat lower than that of TPD/polycarbonate blend at the similar molar ratio, about 62:38, which is attributed to molecular structure confinement of the hole transporting polymer in contrast to molecularly doped polymers like the TPD/polycarbonate blend.  The charge mobility of PABC is nevertheless one of the higher linear chain hole transporting polymers with charge transporting moiety resides in the linear polymer chain.  However large in relative terms, mobilities of the order or $10^{-5}$ cm${}^2$/Vs fall short of the range necessary to produce high fill factors in the solar cells. Using parameters for P3HT:PCBM (1:1), Deibel et al\cite{deibel2008influence}  and Ramirez et al\cite{ramirez2014optimum} demonstrate theoretically the existence of an optimum electron mobility value of about $10^{-1}$ cm${}^2$/Vs to 1 cm${}^2$/Vs with a logarithmic degradation of PEC as mobility moves away from the optimum value.  When the hole mobility falls from 1 cm${}^2$/Vs to $10^{-5}$ cm${}^2$/Vs, PEC drops by about factor of ten. This could be part of the explanation of the low PEC of the PABC:PCBM solar cell. 

\begin{figure}[ht]
  \begin{tabular}{ccc}
    \includegraphics[width=0.8\columnwidth]{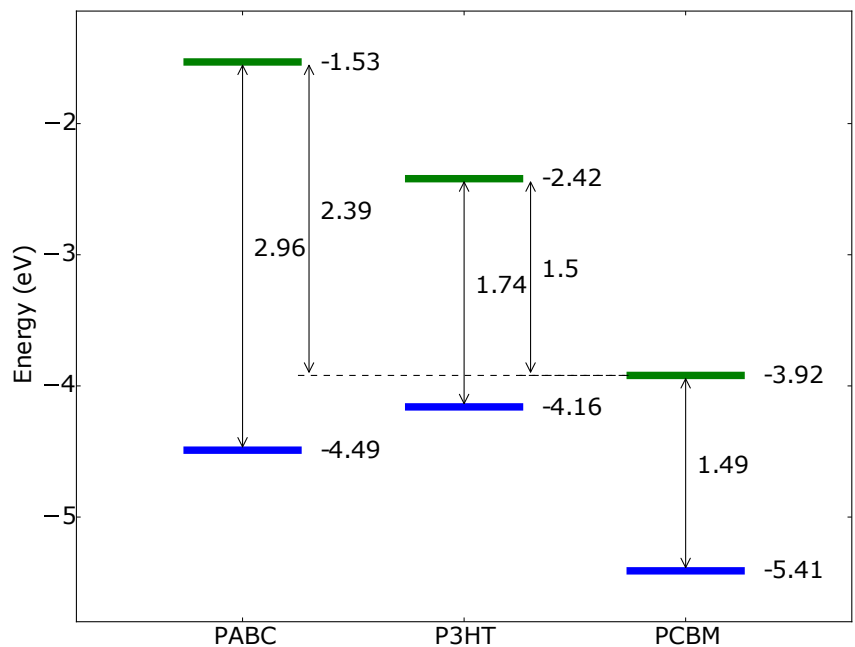}
  \end{tabular}
  \caption{Energy levels for frontier orbitals of PABC, P3HT and PCBM, in the chosen oligomer approximations. LUMO levels are in green, and HOMO levels in blue.}
  \label{energy-levels}
  \end{figure}

The optimum  experimentally found blend ratio for PABC: PCBM is 1 : 3 while the optimum  blend for P3HT:PCBM is approximately\cite{dennler2009polymer} 
1 : 1. This might be explained by the difference in relative mobility of electrons and holes in the blend. The molecular mass of the repeating unit in PABC is 390, where 242 is attributed to arylamine, hole transporting moiety, and 148 to PABC's biscarbonate ester component.  The molecular mass of PCBM is 911, so the molar ratio for the optimum blend ratio of PABC:PCBM is $3*390/911$ or about 1.3 : 1.  This contrasts with the 5.5 : 1 molar ratio for the optimum blend for the P3HT:PCBM system \cite{dennler2009polymer}.  The PABC:PCBM system requires a significantly smaller amount of hole transporting units to reach a balance in electron to hole mobility.   It is well known that the highest power conversion efficiency is reached when there is a balance in the electron to hole mobility ratio\cite{nakamura2005relation,ramirez2014optimum}.
Low mobility combined with steady currents frequently lead to space-charge limited current (SCLC)\cite{zhang2012theory}. In SCLC trapped charges accumulate in the bulk of the device creating a field that reduces throughput. However, observation of a Log-Log plot of the I-V curve for a PABC:PCBM in the dark suggested that in our devices  the traps are shallow and should not be the main efficiency limiting mechanism.
When comparing our devices' J-V curve with that of other materials, as compiled by Zhou \textit{et al}\cite{zhou2012rational}{}, we see that while PABC grants a typical value for $V_{OC}$, it falls an order of magnitude short on the aspect of $J_{SC}$, which we recognize as the limiting factor on PCE. Despite the fact that $J_{SC}$ is essentially a macroscopic observable heavily dependent on the morphology, we decided to apply electronic structure methods  to clarify some of its characteristics. In fact, electronic structure studies are regularly used in the context of organic photovoltaics to link variables of very different scales. As one example, Fu \textit{et al}\cite{fu2014structure}  have calculated the transfer integrals describing exciton dissociation rates in a model OPV bilayer. These dissociation rates can be readily measured and are the main source of photogenerated charge carriers.
As summarized by Li\cite{li2012molecular}, it is accepted that $V_{OC}$ is proportional to the gap between the donor HOMO and acceptor LUMO. To investigate it, we have calculated the frontier levels of the PABC and PCBM molecules, as well as P3HT as a standard for comparison. For computational reasons, in this work we chose to approximate the PABC polymer by a three-unit long olygomer, and P3HT as a six-unit long olygomer. The results can be seen in Fig~\ref{energy-levels}. 

\begin{figure}[ht]
  \begin{tabular}{ccc}
    \includegraphics[width=0.7\columnwidth]{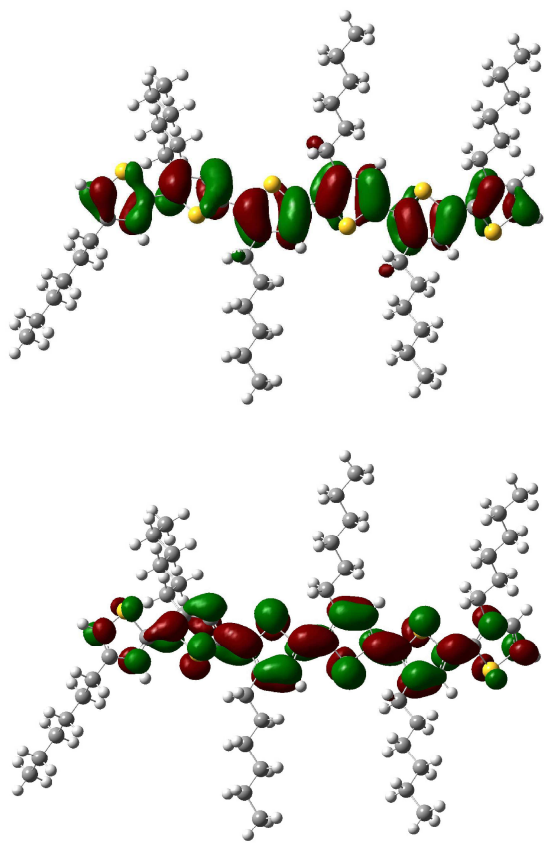}
  \end{tabular}
  \caption{a) HOMO (Left figure) and b) LUMO  orbitals for P3HT oligomer.}
  \label{p3ht-orbitals}
  \end{figure}

As we can see, the PABC:PCBM blend has a larger gap than the P3HT:PCBM blend, which supports our experimental result of typical $V_{OC}$ for our material. This is consistent with our predicted difference between donor and acceptor LUMO, which is larger than the  $0.3$ eV that the literature  usually recognizes as being necessary for good charge separation (see for example Zhou \textit{et al}\cite{zhou2012rational}).
To locate the origin of the limitation of $J_{SC}$, we plotted the frontier orbitals for a series of increasingly long PABC oligomers, and compared them to the frontier orbitals of a six-unit long P3HT oligomer. It can be seen that while the frontier orbitals for P3HT spread along the full span of the polymer, following the $\pi-$bond backbone as typical of most polymer organic semiconductors, Fig~\ref{p3ht-orbitals}, the frontier orbitals of the PABC polymer form a slightly more complicated picture.

\begin{figure}[ht]
  \begin{tabular}{ccc}
    \includegraphics[width=0.7\columnwidth]{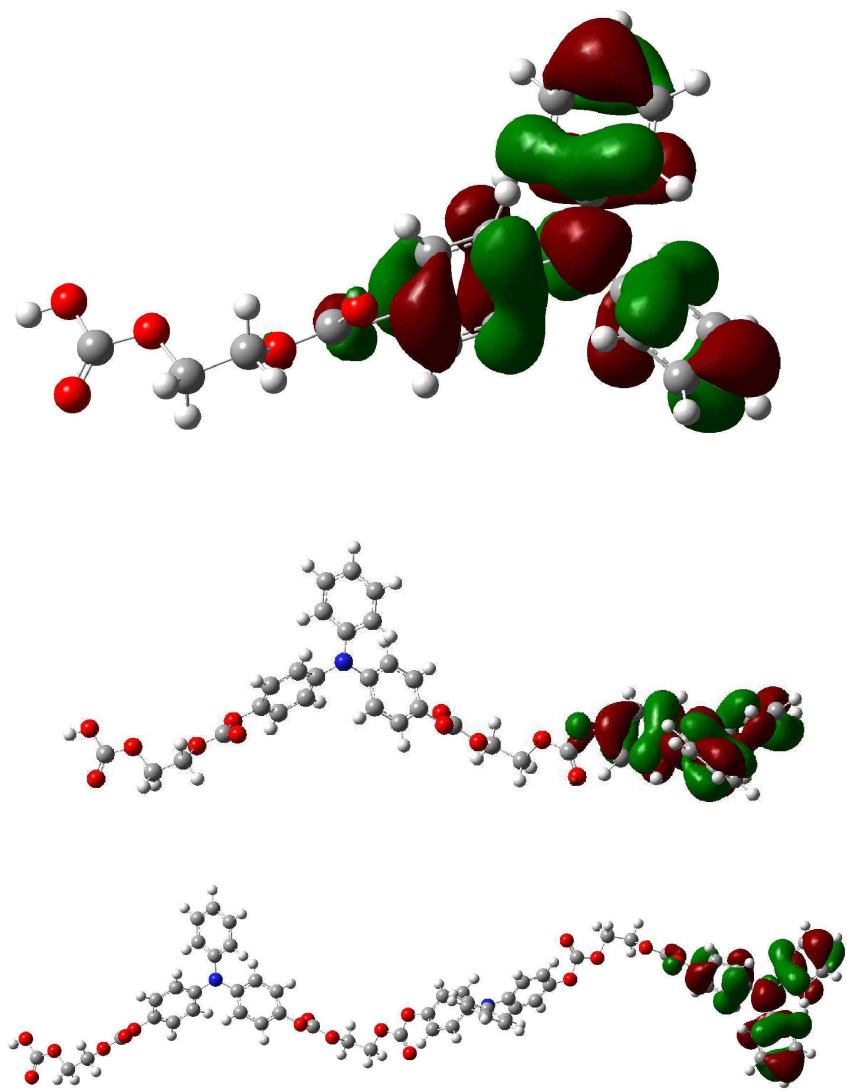}
  \end{tabular}
  \caption{{\small HOMO orbital for PABC oligomers}}
  \label{pabc-homo}
  \end{figure}

\begin{figure}[ht]
  \begin{tabular}{ccc}
    \includegraphics[width=0.7\columnwidth]{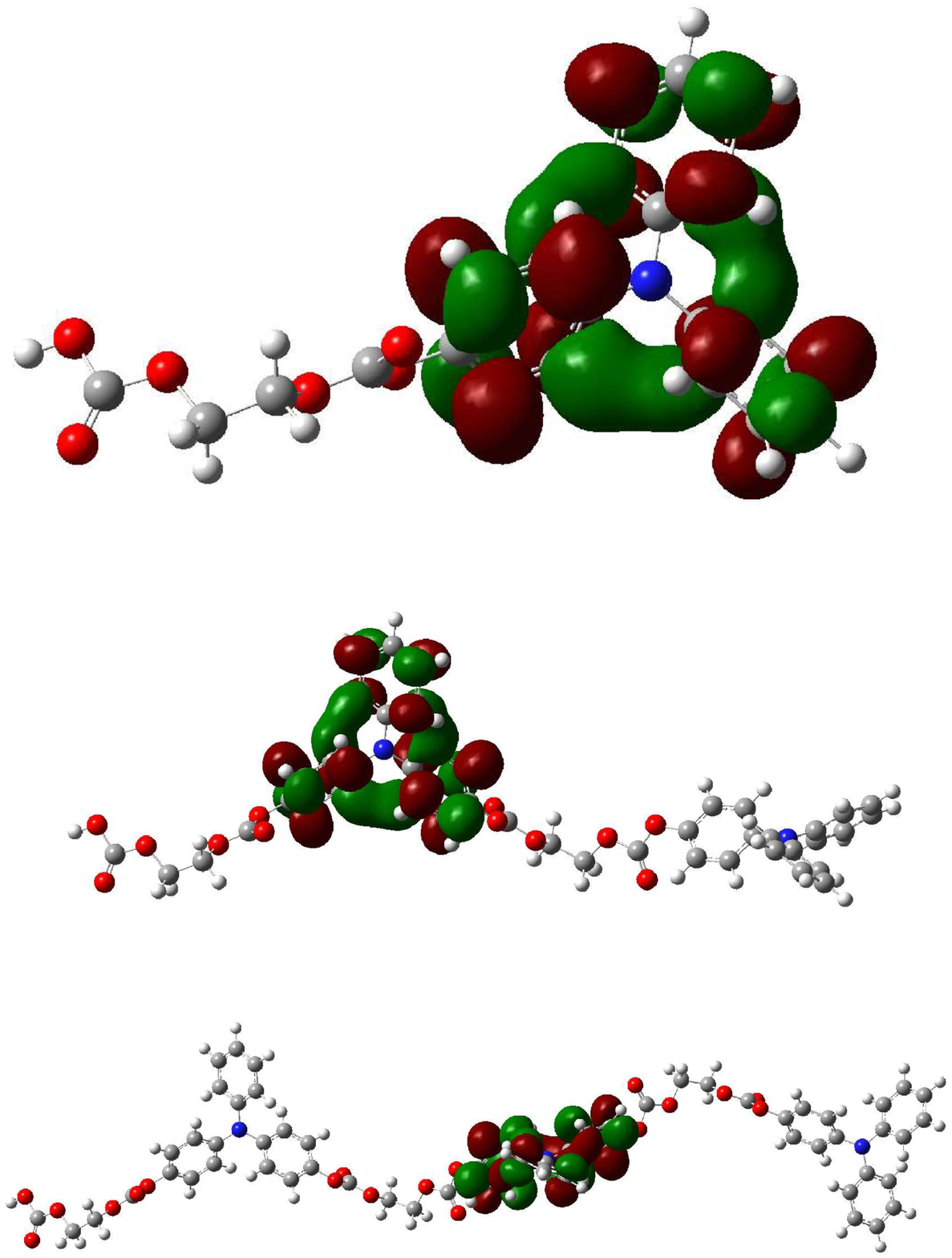}
  \end{tabular}
  \caption{{\small HOMO orbital for PABC oligomers}}
  \label{pabc-lumo}
  \end{figure}

The highest occupied molecular orbitals form a degenerate set of orbitals localized on individual monomer sites. In fact, these frontier orbitals present almost no deformation by the presence of neighbours; they have the same shape as the one calculated for the isolated monomer, see Fig~\ref{pabc-homo}. The same is true for the lowest unoccupied molecular orbitals, see Fig~\ref{pabc-lumo}.

However, degeneracy implies that a linear combination over each set of frontier orbitals can also be understood as a real orbital. Therefore, one can construct a frontier orbital delocalized over the donor polymer sites, which provides a mechanism for charge transport. Nonetheless, even these resulting orbitals, localized over all donor sites, are more localized than those of the P3HT HOMO and LUMO, as the later is fully delocalized over the entire conjugated backbone. The localization observed in PABC will result in less mobile excitations in this polymer when compared to the equivalent excitations in P3HT, and hence this effect could be seen as the hindering factor for excitation transfer between PABC molecules, limiting the charge transport and being the most likely cause for the low $J_{SC}$ observed in experiment.
The frontier orbital DFT calculations were done following geometry optimization, using the B97D functional\cite{grimme2006semiempirical}, with basis 6-311g**. All calculations were done using Gaussian 09\cite{g09} software package.
Efficiency could have an incremental improvement with better ITO conductivity. A P-type oxide semiconductor such as NiO could be used to coat the ITO to increase efficiency by appropriately matching the energy levels\cite{irwin2008p}. Analogously, an n-type semiconductor such as LiF can be used to coat the Al electrode to increase efficiency\cite{hains2007bulk}.  Another opportunity is the optimization of device thickness that is known to increase efficiency both due to recombination and to optical interference issues\cite{hoppe2007inverse,van2009relation}. Strict control of the aluminum evaporation process is also relevant for the overall efficiency of the sample. If aluminum evaporation occurs faster than a critical rate, the combination of heat capacity and heat dissipation of the sample glass is not enough to keep the sample below  200$^{\circ}$C where thermal damage to the polymer starts to occur. All these factors are also related to the optimization of the morphology of the blend\cite{van2010p3ht} that is an art by itself and requires empirical design and trial and error. 

\section*{Conclusion}

In summary, photovoltaic response was observed with a hole transporting polymer polyarylamine biscarbonate ester (PABC) blended with PCBM as the electron transporting material, in the bulk heterojunction configuration.  Efficiencies without further optimization were of the order of 0.5\%.  Our results suggested that the low efficiencies are due to low PCBM mobility for the technique used. 

The experimental part of this work has been disclosed at US Patent 8,975,511 by LBL, GCC, AEP and KB.  DASF acknowledges the financial support from the Brazilian Research Council CNPq, grants 306968/2013-4 and 407682/2013-9, and FAP-DF grant 0193.001.062/2015. AAM and G.C.C. acknowledge  financial support from the Brazilian  Federal Agency  for  Support  and  Evaluation  of  Graduate Education (CAPES)  within  the  Ministry  of Education  of Brazil and CAPES/Programa Ciencia sem Fronteiras.


\end{document}